\documentclass[9pt,conference]{IEEEtran}
\usepackage{amssymb,amsthm,amsmath,array}
\usepackage{graphicx}
\usepackage[caption=false,font=footnotesize]{subfig}
\usepackage{xspace}
\usepackage[sort&compress, numbers]{natbib}
\usepackage{stmaryrd}
\usepackage{xcolor}
\usepackage{mathtools}
\usepackage{float}
\usepackage{textcomp}
\usepackage{amsmath}
\usepackage{amssymb,amsthm,amsmath,array}
\usepackage{graphicx}
\usepackage[caption=false,font=footnotesize]{subfig}
\usepackage{xspace}
\usepackage[sort&compress, numbers]{natbib}
\usepackage{stmaryrd}
\usepackage{xcolor}
\usepackage{mathtools}
\usepackage{float}
\usepackage{textcomp}

\usepackage{standalone}
\graphicspath{{Figures/}}

\usepackage{tikz}
\usetikzlibrary{external}
\usetikzlibrary{arrows}
\usepackage{pgfplots}
\pgfplotsset{compat=1.7}

\newcommand{\bs}{\boldsymbol}
\newcommand{\bb}{\mathbb}
\newcommand{\cl}{\mathcal}

\newcommand{\ie}{\emph{i.e.},\xspace}

\newcommand{\sq}{\vspace{-2mm}}

\begin{document}
\title{\textit{In silico} Ptychography of Lithium-ion Cathode Materials from Subsampled 4-D STEM Data}
\author{\IEEEauthorblockN{
        Alex W. Robinson\IEEEauthorrefmark{1},
        Amirafshar Moshtaghpour\IEEEauthorrefmark{1}\IEEEauthorrefmark{3},
        Jack Wells\IEEEauthorrefmark{2},
        Daniel Nicholls\IEEEauthorrefmark{1},
        Zoë Broad\IEEEauthorrefmark{1}, \\
        Angus I. Kirkland\IEEEauthorrefmark{3}\IEEEauthorrefmark{4},
        Beata L. Mehdi\IEEEauthorrefmark{1},
        Nigel D. Browning\IEEEauthorrefmark{1}\IEEEauthorrefmark{5}\IEEEauthorrefmark{6}
    }
    \IEEEauthorblockA{
        \IEEEauthorrefmark{1} Department of Mechanical, Materials and Aerospace Engineering, University of Liverpool, UK.\\
        \IEEEauthorrefmark{2} Distributed Algorithms Centre for Doctoral Training, University of Liverpool, Liverpool, UK.\\
        \IEEEauthorrefmark{3} Correlated Imaging Group, Rosalind Franklin Institute, Harwell Science and Innovation Campus, Didcot, UK.\\
        \IEEEauthorrefmark{4} Department of Materials, University of Oxford, UK.\\
        \IEEEauthorrefmark{5} Physical and Computational Science Directorate, Pacific Northwest National Laboratory, Richland, USA.\\
        \IEEEauthorrefmark{6} Sivananthan Laboratories, 590 Territorial Drive, Bolingbrook, IL, USA.}
}
\maketitle
\begin{abstract}
    High quality scanning transmission electron microscopy (STEM) data acquisition and analysis has become increasingly important due to the commercial demand for investigating the properties of complex materials such as battery cathodes; however, multidimensional techniques (such as $4$-D STEM) which can improve resolution and sample information are ultimately limited by the beam-damage properties of the materials or the signal-to-noise ratio of the result. subsampling offers a solution to this problem by retaining high signal, but distributing the dose across the sample such that the damage can be reduced. It is for these reasons that we propose a method of  subsampling for $4$-D STEM, which can take advantage of the redundancy within said data to recover functionally identical results to the ground truth. We apply these ideas to a simulated $4$-D STEM data set of a LiMnO$_{2}$ sample and we obtained high quality reconstruction of phase images using 12.5\% subsampling. 
\end{abstract}
\section{Introduction}
\begin{figure}[t!]
    \centering
    \scalebox{0.5}{\includegraphics{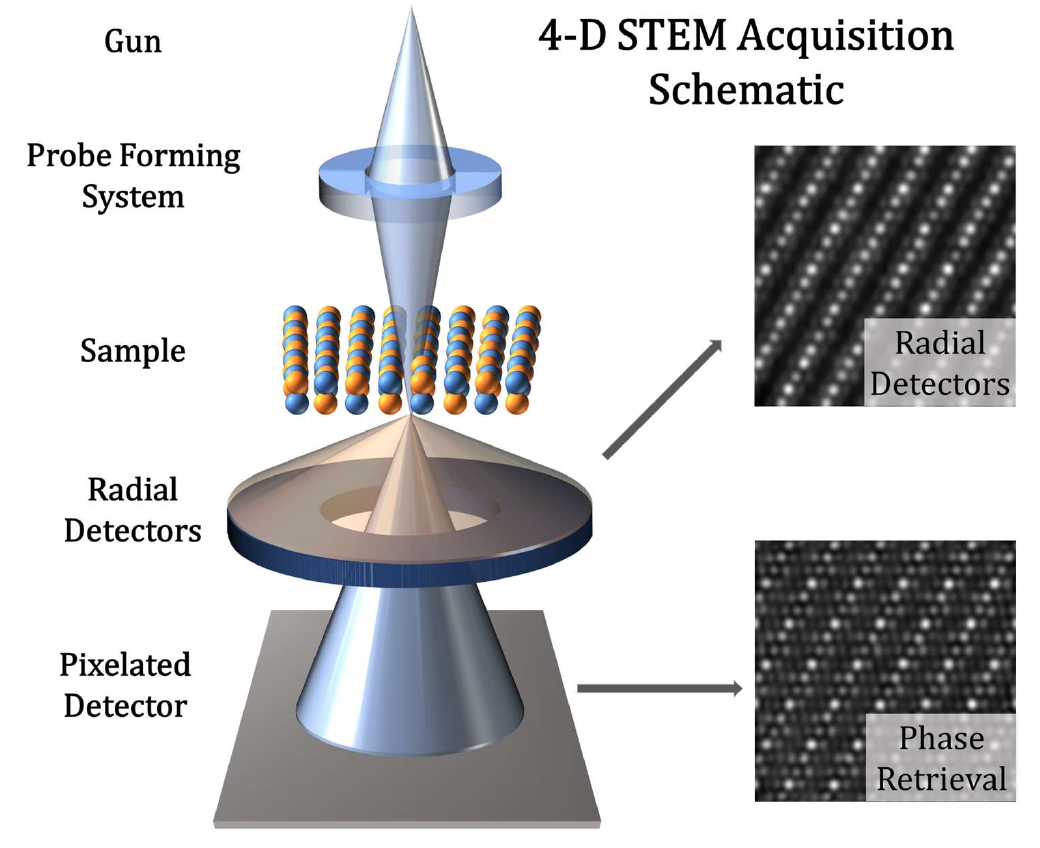}}
    \sq
    \caption{\textbf{Schematic of 4-D STEM acquisition.} A focused electron beam is scanned over the sample in a raster fashion and a convergent beam diffraction pattern is acquired on a pixelated detector. Fixed radial detectors can also be used simultaneously to acquire, for example, a Z-contrast image.}
    \label{fig:my_label}
    \sq\
\end{figure}

Scanning transmission electron microscopy (STEM) is a powerful tool for imaging the complex sub-nanoscale structures of various materials, such as battery materials, or nano-electronics. In STEM, a focused beam of electrons is moved over the sample in a raster fashion and various signals are acquired. Four-dimensional STEM ($4$-D STEM) is growing within the field thanks to its multi-modal imaging methods. As shown in Fig.~\ref{fig:my_label}, a $2$-dimensional (2-D) diffraction pattern is acquired by a pixelated detector for each probe location, hence forming a total data set where each data-point is indexed by its reciprocal and real space locations~\cite{ophus2019four}. First developed by Hoppe in 1969~\cite{hoppe1969beugung}, one popular use case of $4$-D STEM is in ptychography -- the phase change induced by the sample is recovered through a phase retrieval algorithm, meaning low atomic number elements can be resolved~\cite{yang2017electron}.

However, the advances made over recent years to reduce lens aberrations~\cite{krivanek1999towards} have led to smaller, more intense electron probes~\cite{batson2002sub}, which can ultimately destroy the sample prior to reliable observation. Furthermore, if the user wishes to collect 4-D STEM data as demonstrated in Fig.~\ref{fig:my_label}, this can increase acquisition times and lead to further exposure of the beam to the sample. A common approach to overcoming the limitations of STEM for beam-sensitive materials is through low-dose techniques~\cite{bustillo20214d, li20224d,zhou2020low}. However, this comes with lower signal-to-noise ratio (SNR)~\cite{pennycook2019high}.  
An alternative method to overcome the issues of acquisition speed and beam damage is through subsampling. This has been demonstrated as a powerful technique for this purpose in several works, including annular dark-field (ADF) STEM imaging~\cite{nicholls2022compressive}, focused ion beam (FIB) microscopy with scanning electron microscopy (SEM)~\cite{nicholls2022targeted}, $4$-D STEM~\cite{stevens2018subsampled}, and STEM simulations~\cite{robinson2023towards,robinson2022sim}. 

This work aims to show that $4$-D STEM subsampling is robust to the recovery and observation of lithium in the case of ptychography, by first subsampling and inpainting the data, followed by applying the Wigner Distribution Deconvolution (WDD) algorithm to recover phase images.

\section{Methods}
In this section we present our compressive $4$-D STEM acquisition model, data recovery method, and phase retrieval solution using WDD. 

\vspace{1mm}
\noindent\textbf{Acquisition model.} 
Let $\cl X\in \bb R^{P_{1}\times P_{2}\times D_{1} \times D_{2}}$ be the discretised $4$-dimensional representation of fully sampled 4-D STEM data using an electron probe scanning $P_1\times P_2$ positions on the sample and using a $D_1 \times D_2$ pixelated detector. This data can be represented as a matrix $\bs X \in \bb R^{P \times D}$, with $P = P_1 P_2$ rows and $D = D_1 D_2$ rows: each row contains a vectorised version of a diffraction pattern collected at one probe location and each column contains a vectorised version of a probe-wise slice of the 4-D data collected at one detector pixel.
We now introduce our compressed $4$-D STEM system to reduce beam damage and increase acquisition speed. We do this by subsampling $M \ll P$ probe positions collected in a subsampling set $\Omega\subset\{1, \cdots, P\}$ with cardinality $\lvert\Omega\rvert = M$. Mathematically, for every $d^{\rm th}$ column of $\bs X$, \ie $\bs x_d \in \bb R^{P}$, 
\begin{equation}\label{eq:cseq}
\bs{y}_{d} := \bs{P}_{\Omega}\bs{x}_d + \bs{n}_d \in \bb R^{P},
\end{equation} 
where $\bs{P}_{\Omega}\in\{{0,1}\}^{P\times P}$ is a mask operator with $(\bs{P}_{\Omega}x)_{j} = \bs{x}_j$ if $j\in\Omega$ (and zero, otherwise), and $\bs{n}_d$ is a noise.
\vspace{1mm}

\noindent\textbf{Data recovery method.} We now consider the recovery of $\hat{\bs x}_d\approx\bs{x}_d$ from subsampled measurements $\bs{y}_{d}$ in \eqref{eq:cseq}, \ie an inpainting problem. We assume that each slice is sparse in some unknown dictionary, which can be learnt during the recovery process. We accordingly choose the Beta Process Factor Analysis (BPFA) method~\cite{paisley2014bayesian,sertoglu2015scalable} to recover both the dictionary and the target signal $\bs{\hat{x}}_d$. We note that for each slice a dictionary is learnt independently and the case of inpainting the whole 4-D data is left for a future work.

\vspace{1mm}
\noindent\textbf{Phase retrieval.} In order to recover the phase of the $4$-D STEM data, we chose to use the WDD algorithm~\cite{rodenburg1992theory}. The method has been well established as a suitable method to recover object phase at focused probe conditions~\cite{martinez2017comparison,o2021contrast}. The goal of the algorithm is to estimate the object function given diffraction patterns and an estimate of the initial probe. Initially, a Fourier transform of the data set with respect to the probe locations is performed, followed by an inverse Fourier transform with respect to the reciprocal space coordinates. The result follows the mathematical definition of a Wigner distribution function, so can be deconvolved through a Wiener deconvolution. Following application of the Wiener deconvolution, we apply a Fourier transform with respect to the reciprocal space coordinates.
By doing this, small phase shifts can be measured, and hence the phase of the object is retrieved.  

\section{Results}
Here we present the simulation set-up and results of the methods described above to a simulated 4-D STEM data set of lithium manganese oxide sample (LiMnO$_{2}$) using the abTEM software package which can be found at \cite{madsen2021abtem}. 

\begin{figure*}[t]
\begin{minipage}{\textwidth}
    \centering
    \includegraphics[width=0.8\textwidth]{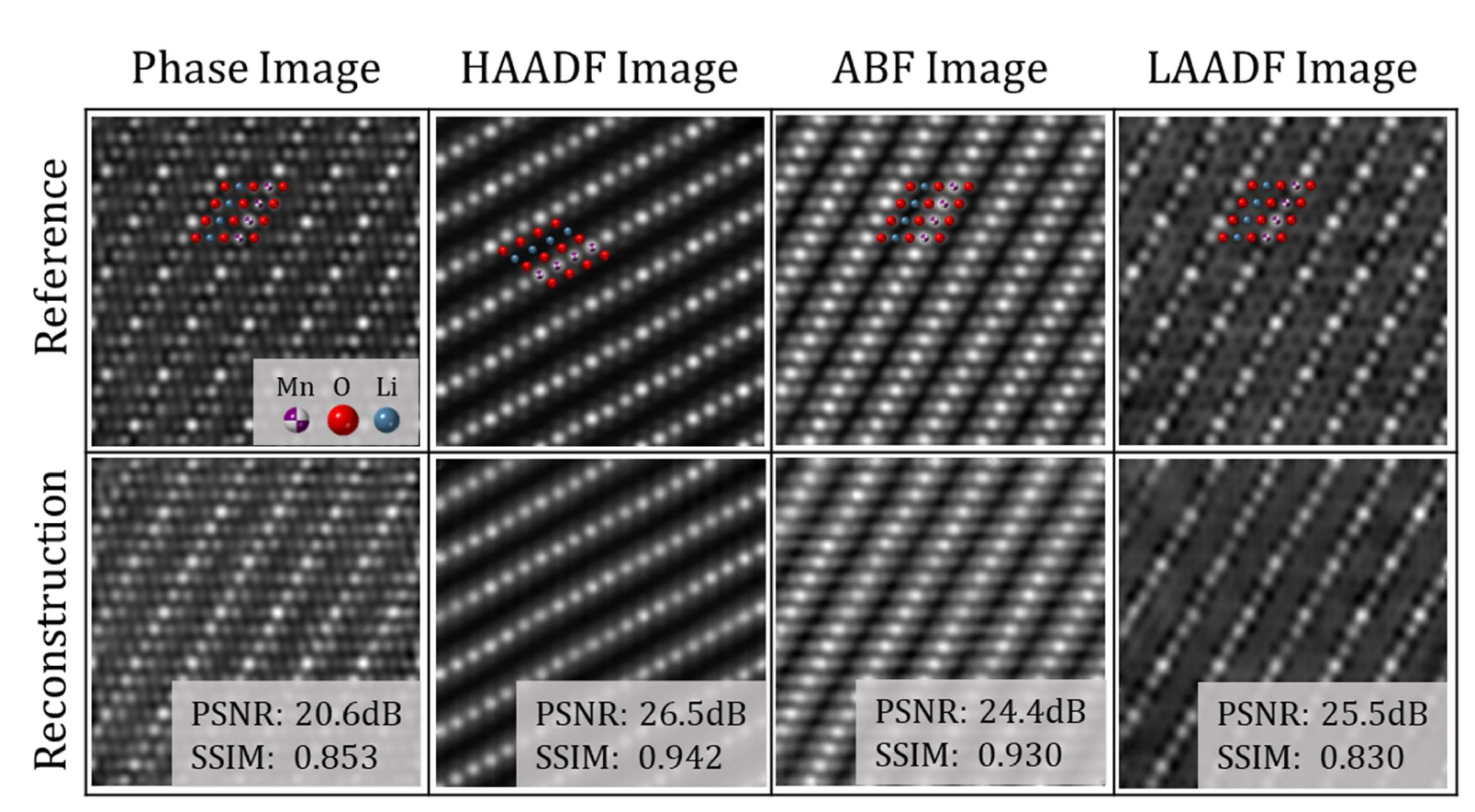}
    \caption{\textbf{Results of subsampling and inpainting.} The HAADF and phase images can be suitably recovered at $12.5\%$ probe subsampling. Furthermore, we achieve high quality recovery of other signals from our 4-D data, yet qualitative observation of lithium is only identifiable in the phase image.}
    \label{fig:result}
    \end{minipage}
    \sq\sq
\end{figure*}

\vspace{1mm}
\noindent\textbf{Simulation set-up.} Through simulations, we obtained a 4-D data set of convergent beam diffraction pattern at each probe location, as well as a high angle ADF (HAADF) image, low angle ADF (LAADF) image, and annular bright field (ABF) image (see in Fig.~\ref{fig:result}).
The simulation was performed using a $200$kV electron beam and a $25$mrad convergence semi-angle. The defocus value was set to $1$nm, and a spherical aberration coefficient of $1\mu$m was used. The sample has a thickness of $4.8$nm, and was oriented in the $[101]$ plane to see the layered structure. The inner and outer angles of the HAADF detector were $80$ and $150$ mrad respectively, and the total 4D-STEM data set had dimensions $(D_1,D_2,P_1,P_2) = (81,81,128,128)$. We chose a probe subsampling ratio of $12.5\%$ for both the HAADF and 4D-STEM data.

\vspace{1mm}
\noindent\textbf{Results.} The HAADF and phase images have been suitably recovered from a subsampled data set. The peak signal-to-noise (PSNR) and structural similarity index measure (SSIM) values are indicative of high quality recovery in Fig.~\ref{fig:result}. The phase images were standardised to avoid inconsistencies through linear phase shifts, and it can be seen on the image that a background exists which is not in the reference. This could be an artefact from the inpainting method, however all necessary information is recovered regardless. This artefact could potentially be resolved through parameter optimisation, given the entire data set was recovered with equal parameters for all slices.

When comparing the Z-contrast images and phase reconstructions, the lithium column is visible in the latter but not in the former. Furthermore, the same observation can be seen in both the reference and reconstructed images. This shows that under ideal conditions subsampling can be employed when attempting to acquire $4$-D STEM data of lithium-ion cathode materials.

\section{Conclusion}
We demonstrated the proof-of-concept of subsampling to the ptychographic reconstruction of battery materials. The real challenge is to employ these methods into practice.
In the presentation, realistic noise conditions, drift, and scanning limitations shall be considered to emulate STEM conditions, as well as considering a constrained dose-budget. Furthermore, a range of probe subsampling ratios will be demonstrated. It is also important to note that these methods could lead the way towards 4-D STEM video, which would provide an invaluable insight into the degradation mechanisms of Li-ion batteries.

\bibliographystyle{IEEEtran}
\bibliography{references}

\begin{thebibliography}{10}
\providecommand{\url}[1]{#1}
\csname url@samestyle\endcsname
\providecommand{\newblock}{\relax}
\providecommand{\bibinfo}[2]{#2}
\providecommand{\BIBentrySTDinterwordspacing}{\spaceskip=0pt\relax}
\providecommand{\BIBentryALTinterwordstretchfactor}{4}
\providecommand{\BIBentryALTinterwordspacing}{\spaceskip=\fontdimen2\font plus
\BIBentryALTinterwordstretchfactor\fontdimen3\font minus
  \fontdimen4\font\relax}
\providecommand{\BIBforeignlanguage}[2]{{%
\expandafter\ifx\csname l@#1\endcsname\relax
\typeout{** WARNING: IEEEtran.bst: No hyphenation pattern has been}%
\typeout{** loaded for the language `#1'. Using the pattern for}%
\typeout{** the default language instead.}%
\else
\language=\csname l@#1\endcsname
\fi
#2}}
\providecommand{\BIBdecl}{\relax}
\BIBdecl

\bibitem{ophus2019four}
C.~Ophus, ``Four-dimensional scanning transmission electron microscopy
  (4d-stem): From scanning nanodiffraction to ptychography and beyond,''
  \emph{Microscopy and Microanalysis}, vol.~25, no.~3, pp. 563--582, 2019.

\bibitem{hoppe1969beugung}
W.~Hoppe, ``Beugung im inhomogenen prim{\"a}rstrahlwellenfeld. i. prinzip einer
  phasenmessung von elektronenbeungungsinterferenzen,'' \emph{Acta
  Crystallographica Section A: Crystal Physics, Diffraction, Theoretical and
  General Crystallography}, vol.~25, no.~4, pp. 495--501, 1969.

\bibitem{yang2017electron}
H.~Yang, I.~MacLaren, L.~Jones, G.~T. Martinez, M.~Simson, M.~Huth, H.~Ryll,
  H.~Soltau, R.~Sagawa, Y.~Kondo \emph{et~al.}, ``Electron ptychographic phase
  imaging of light elements in crystalline materials using wigner distribution
  deconvolution,'' \emph{Ultramicroscopy}, vol. 180, pp. 173--179, 2017.

\bibitem{krivanek1999towards}
O.~Krivanek, N.~Dellby, and A.~Lupini, ``Towards sub-{\aa} electron beams,''
  \emph{Ultramicroscopy}, vol.~78, no. 1-4, pp. 1--11, 1999.

\bibitem{batson2002sub}
P.~E. Batson, N.~Dellby, and O.~L. Krivanek, ``Sub-{\aa}ngstrom resolution
  using aberration corrected electron optics,'' \emph{Nature}, vol. 418, no.
  6898, pp. 617--620, 2002.

\bibitem{bustillo20214d}
K.~C. Bustillo, S.~E. Zeltmann, M.~Chen, J.~Donohue, J.~Ciston, C.~Ophus, and
  A.~M. Minor, ``4d-stem of beam-sensitive materials,'' \emph{Accounts of
  chemical research}, vol.~54, no.~11, pp. 2543--2551, 2021.

\bibitem{li20224d}
G.~Li, H.~Zhang, and Y.~Han, ``4d-stem ptychography for electron-beam-sensitive
  materials,'' \emph{ACS Central Science}, 2022.

\bibitem{zhou2020low}
L.~Zhou, J.~Song, J.~S. Kim, X.~Pei, C.~Huang, M.~Boyce, L.~Mendon{\c{c}}a,
  D.~Clare, A.~Siebert, C.~S. Allen \emph{et~al.}, ``Low-dose phase retrieval
  of biological specimens using cryo-electron ptychography,'' \emph{Nature
  communications}, vol.~11, no.~1, pp. 1--9, 2020.

\bibitem{pennycook2019high}
T.~J. Pennycook, G.~T. Martinez, P.~D. Nellist, and J.~C. Meyer, ``High dose
  efficiency atomic resolution imaging via electron ptychography,''
  \emph{Ultramicroscopy}, vol. 196, pp. 131--135, 2019.

\bibitem{nicholls2022compressive}
D.~Nicholls, A.~Robinson, J.~Wells, A.~Moshtaghpour, M.~Bahri, A.~Kirkland, and
  N.~Browning, ``Compressive scanning transmission electron microscopy,'' in
  \emph{Proceedings of the IEEE International Conference on Acoustics, Speech
  and Signal Processing (ICASSP)}, 2022, pp. 1586--1590.

\bibitem{nicholls2022targeted}
D.~Nicholls, J.~Wells, A.~W. Robinson, A.~Moshtaghpour, M.~Kobylynska, R.~A.
  Fleck, A.~I. Kirkland, and N.~D. Browning, ``A targeted sampling strategy for
  compressive cryo focused ion beam scanning electron microscopy,'' \emph{arXiv
  preprint arXiv:2211.03494}, 2022.

\bibitem{stevens2018subsampled}
A.~Stevens, H.~Yang, W.~Hao, L.~Jones, C.~Ophus, P.~D. Nellist, and N.~D.
  Browning, ``Subsampled stem-ptychography,'' \emph{Applied Physics Letters},
  vol. 113, no.~3, p. 033104, 2018.

\bibitem{robinson2023towards}
A.~W. Robinson, J.~Wells, D.~Nicholls, A.~Moshtaghpour, M.~Chi, A.~I. Kirkland,
  and N.~D. Browning, ``Towards real-time stem simulations through targeted
  sub-sampling strategies,'' \emph{Journal of microscopy}, 2023.

\bibitem{robinson2022sim}
A.~W. Robinson, D.~Nicholls, J.~Wells, A.~Moshtaghpour, A.~Kirkland, and N.~D.
  Browning, ``Sim-stem lab: Incorporating compressed sensing theory for fast
  stem simulation,'' \emph{Ultramicroscopy}, vol. 242, p. 113625, 2022.

\bibitem{paisley2014bayesian}
J.~W. Paisley, D.~M. Blei, and M.~I. Jordan, ``Bayesian nonnegative matrix
  factorization with stochastic variational inference.'' 2014.

\bibitem{sertoglu2015scalable}
S.~Sertoglu and J.~Paisley, ``Scalable {B}ayesian nonparametric dictionary
  learning,'' in \emph{2015 23rd European Signal Processing Conference
  (EUSIPCO)}, 2015, pp. 2771--2775.

\bibitem{rodenburg1992theory}
J.~Rodenburg and R.~Bates, ``The theory of super-resolution electron microscopy
  via wigner-distribution deconvolution,'' \emph{Philosophical Transactions of
  the Royal Society of London. Series A: Physical and Engineering Sciences},
  vol. 339, no. 1655, pp. 521--553, 1992.

\bibitem{martinez2017comparison}
G.~Martinez, M.~Humphry, and P.~Nellist, ``A comparison of phase-retrieval
  algorithms for focused-probe electron ptychography,'' \emph{Microscopy and
  Microanalysis}, vol.~23, no.~S1, pp. 476--477, 2017.

\bibitem{o2021contrast}
C.~M. O’Leary, G.~T. Martinez, E.~Liberti, M.~J. Humphry, A.~I. Kirkland, and
  P.~D. Nellist, ``Contrast transfer and noise considerations in focused-probe
  electron ptychography,'' \emph{Ultramicroscopy}, vol. 221, p. 113189, 2021.

\bibitem{madsen2021abtem}
J.~Madsen and T.~Susi, ``The abtem code: transmission electron microscopy from
  first principles,'' \emph{Open Research Europe}, vol.~1, no.~24, p.~24, 2021.

\end{thebibliography}
\end{document}